\documentclass[aps,prl,twocolumn,floatfix]{revtex4}
\usepackage{epsfig,amssymb,amsmath}
\newcommand{\be}{\begin{equation}}    \newcommand{\bea}{\begin{eqnarray}}   \newcommand{\ra}{\rangle}
\newcommand{\ee}{\end{equation}}      \newcommand{\eea}{\end{eqnarray}}     \newcommand{\la}{\langle}
\newcommand{\RR} {\rangle\!\rangle}   
\newcommand{\LL}{\langle\!\langle}    
\begin{document}
\title{Detection of non-Gaussian Fluctuations in a Quantum Point Contact}
\author{G. Gershon$^a$, Yu. Bomze$^{a,b}$,
E. V. Sukhorukov$^c$, and M. Reznikov$^a$}
\affiliation{$^a$ Department of Physics and Solid State
Institute,
Technion-IIT, Haifa 32000, Israel\\
$^b$ Physics Department, Duke University, Durham, NC 27708\\
$^c$ D\'{e}partment de Physique Th\'{e}orique,
Universit\'{e} de Gen\`{e}ve, CH-1211 Gen\`{e}ve 4,
Switzerland}

\begin{abstract}
An experimental study of current fluctuations through a
tunable transmission barrier, a quantum point contact, are
reported. We measure the probability distribution function
of transmitted charge with precision sufficient to extract
the first three cumulants. To obtain the intrinsic
quantities, corresponding to voltage-biased barrier, we
employ a procedure that accounts for the response of the
external circuit and the amplifier. The third cumulant,
obtained with a high precision, is found to agree with the
prediction for the statistics of transport in the
non-Poissonian regime.
\end{abstract}
\maketitle

Recently, measurements of current fluctuations arising from
charge discreteness have become an invaluable tool in
mesoscopic physics, the most noticeable achievement being
the shot noise measurement of quasi-particle charge in the
fractional quantum Hall effect \cite{Glattli,dePicciotto}.
Typically, mesoscopic shot noise experiments  report
zero-frequency noise power, but this quantity contains only
partial information about the statistics of the transmitted
charge. The counting statistics (CS) \cite{Levitov93}
is entirely characterized by the set of cumulants
(irreducible moments) $\LL q^n\RR$ of the charge $q(\tau)$
transmitted through a voltage-biased system during a
sampling time $\tau$. In the long time limit they are
proportional to the time, $\LL q^n\RR=\LL J^n\RR\, \tau$;
this expression defines the {\em current cumulants} $\LL
J^n\RR$. For example, Gaussian noise is fully determined by
the first two current cumulants, the average current $I=\LL
J\RR$ and the noise power $S=\LL J^2\RR$. The simplest
measure of the non-Gaussianity, the third current cumulant
$S=\LL J^3\RR$ which reflects the skewness of the current
distribution, is the central focus of our paper. It is
linear and universal at low bias voltage, and therefore may
be used as a tool for investigation of strongly correlated
systems, where large bias cannot be applied without
substantially affecting their properties.

However, during a typical sampling time a large number of
electrons passes through the system. This fact, by virtue
of the central limit theorem, makes it difficult to observe
non-Gaussian effects in electron
transport~\cite{footnote0}, unless electrons are counted
one by one as, e.\,g., in Coulomb blockaded quantum dots
\cite{Fujisawa06,Gustavsson06,Gustavsson06a}. To date, $\LL
J^3\RR$ has been measured only in low transmission
tunneling junctions either by measuring voltage on a load
resistor \cite{ReuletPRL,Bomze05}, or with the help of the
on-chip Josephson threshold detector \cite{Pekola06}.

In a typical experiment, due to the electrons' charge, the
voltage across the sample is not constant, so the measured
statistics are not trivially related to the CS of the
voltage-biased system. Indeed, the original experiment on
the third cumulant of a tunneling current \cite{ReuletPRL}
exhibited totally unexpected results, which were
explained \cite{KNB02,Nagaev02,Beenakker03,Geneva_group} by
the back-action of the measurement apparatus on the sample.
That is, in addition to $\LL J^3\RR$, the experimentally
measured potential fluctuations also contain contributions,
dubbed ``cascade corrections'' in Ref.\,\cite{Nagaev02},
from the voltage dependent current $I$ and noise power $S$.
Surprisingly, even if the load resistance is small, the
cascade corrections may be of the same order as $\LL
J^3\RR$.

\begin{figure}[tpb]
\includegraphics [angle=-90, width=3.375 in] {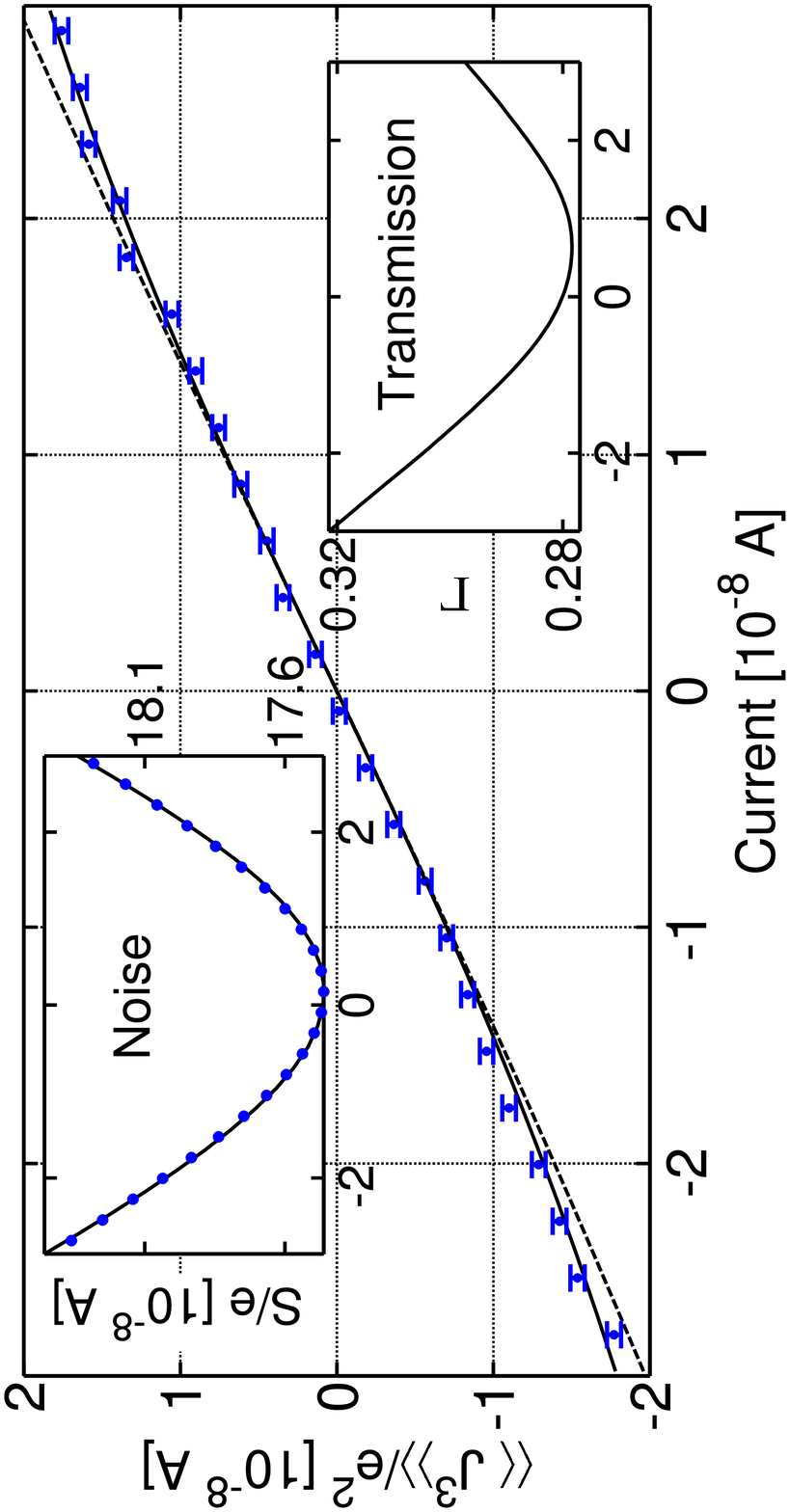} 
\caption{ The current cumulant $\LL J^3\RR$ at
$\Gamma\sim 0.3$, derived from the experimental results for
$\LL V_a^2\RR$ and $\LL V_a^3\RR$ using Eq.\
(\ref{V3}-\ref{Vnl}) for the sample QPC2 measured with
amplifier $b$. Solid line is the theoretical prediction for
$\LL J^3\RR$, Eq.\,(\ref{S3}); the dashed line is the
low-bias limit, Eq.\ (\ref{S3l}), with current-dependent
transmission $\Gamma$ (lower inset). Upper inset: Measured
value of the noise power $S$ versus current; solid line is
the Eq.\,(\ref{S2}).} \vspace{-5mm} \label{S3+S2}
\end{figure}

We report here the first measurement of the third cumulant
of the {\em non-Poissonian} current partitioned by a
variable transmission barrier. To obtain the third
cumulant, we develop a procedure that allows us to separate
the $\LL J^3\RR$ contribution and cascade corrections in a
reliable fashion. In particular, the frequency dependence
of the amplifier gain is found to affect these
contributions differently. The resulting cumulant $\LL
J^3\RR$ agrees accurately with the predictions of
\cite{Levitov93} for all transmissions, $0<\Gamma<1$,
without fitting parameters.
\begin{figure}[tpb]
\includegraphics [angle=-90, width=3.375 in]{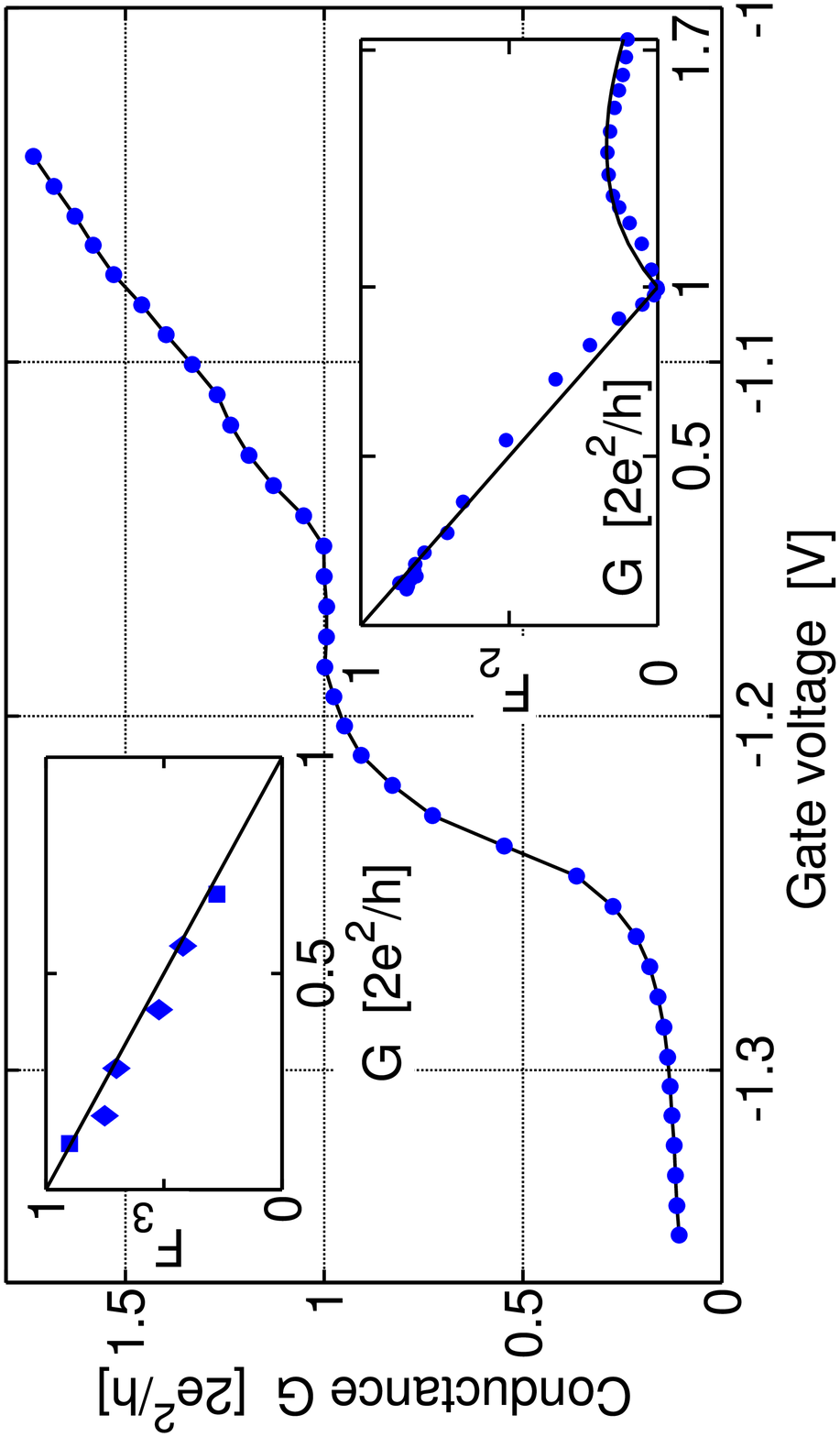}
\caption{The conductance step of $g_0=2e^2/h$ versus gate
voltage $V_g$ used to control the transmission $\Gamma$ of
QPC conduction channels is shown for one of our samples.
Lower inset: The Fano factor $F_2=S/eI$ versus the
QPC conductance $G/g_0$ for the same sample. Solid line is
the theoretical prediction, Eq.\ (\ref{S2}), for the large
bias, $eV\gg k_{\rm B}T$  ($T=1.8\,{\rm K}$, $B=2.2\,{\rm
T}$). Upper inset: The Fano factor $F_3=\LL J^3\RR/e^2I$
versus $\Gamma$ at low bias, $eV\ll k_{\rm B}T$ ($T\approx
5\,{\rm K}$), for QPC1 ($\blacklozenge$), and QPC2
($\blacksquare$). Theoretical prediction $F_3= (1-\Gamma)$
is shown by the solid line. } \label{qpc}
\end{figure}
As a variable transmission barrier we use a quantum point
contact (QPC) -- a small quasi-1D constriction formed in a
2D electron gas by negative voltage $V_g$ applied to split
gates~\cite{qpc}. By varying $V_g$ one can gradually
increase the width of the constriction.
Fig.\ \ref{qpc} shows a typical dependence of the QPC
conductance on $V_g$. The $g_0=2e^2/h$ step in conductance
vs. $V_g$, and the accompanying noise minimum (bottom
inset), defines the lowest spin-degenerate QPC channel used
in this work. We define the average QPC transmission as:
\be \label{Gamma} \Gamma(I)=\frac{I}{g_0 V_s(I)}, \ee
where $I$ is the average current and $V_s$ is the voltage
across the sample. The  observed noise agrees well with the
theoretical expectations\,\cite{BB} and has only thermal
and shot noise contributions. No material contribution,
e.g.\ $1/f$ noise, is detected. The measurements are done
at elevated temperature $T\approx5$\,K in order to reduce
nonlinearity of the current-voltage characteristic.
Nevertheless, for QPC conductance below $2e^2/h$, the
transport is dominated by a single channel, as confirmed by
the agreement of the measured noise with the single-channel
expectations (see Fig.\,\ref{S3+S2}).
\begin{figure}[tbp]
\includegraphics [width=3.0 in] {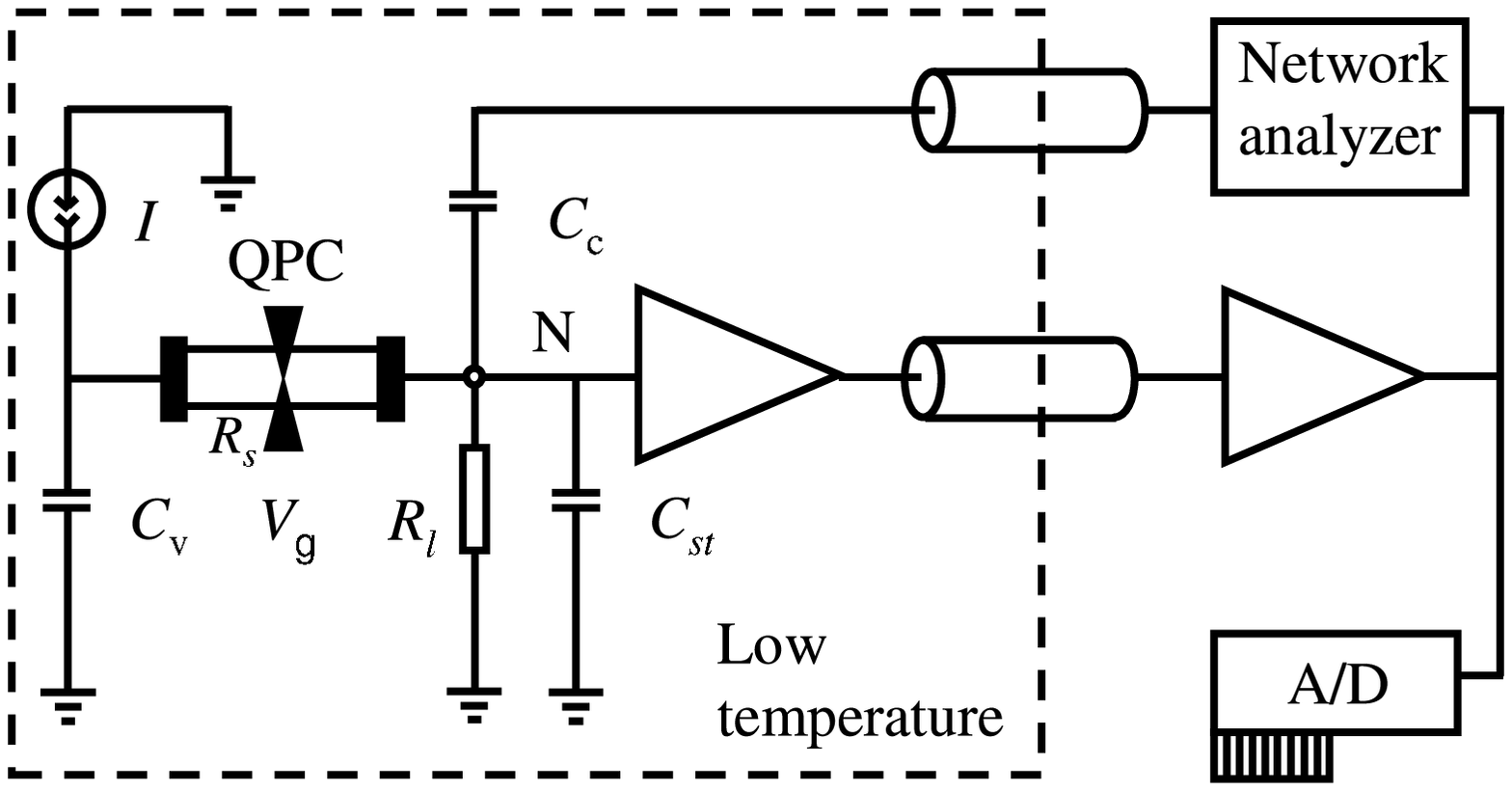}
%\vspace{-4mm}
\caption{ Experimental setup schematic. The current source
drives constant {\it average} current $I$ through the QPC
sample, while the capacitor $C_V$ fixes the voltage across
the sample. The QPC resistance $R_s$ can be tuned by the
gate voltage $V_g$. $R_{l}$ is the load resistance and
$C_{st}$ is the stray capacitance  of the wires at the
amplifier input (the point N).  Current fluctuations
generate voltage fluctuations at the point $N$, which are
amplified and digitized by 12 bit analog-to-digital
convertor (A/D). We introduce current through the capacitor
$C_c$ and measure the response with a network analyzer to
calibrate the setup.} \vspace{-5mm} \label{Setup}
\end{figure}

It is instructive to discuss the issues encountered in the
investigation of the third cumulant prior to presenting our
experimental results. We focus on the experimental setup
shown in Fig.\ \ref{Setup}, however our conclusions are
quite general. Current fluctuations $J$ generated by the
QPC sample and by thermal noise in the load resistor $R_l$
give rise to the voltage $V_N(\omega)=Z(\omega)J(\omega)$
at the amplifier input (point N in Fig.\ \ref{Setup}),
where the impedance
$Z(\omega)=R_\parallel/(1-i\omega\tau_{RC})$ is determined
by $R_\parallel=R_{s}R_l/(R_{s}+R_l)\approx 7\,{\rm KOhm}$,
and by the time constant $\tau_{RC}=R_\parallel
C_\parallel$ (here $C_\parallel=C_{st}+C_c\approx 3\,{\rm
pf}$) which sets the high frequency cut-off of the circuit
$1/(2 \pi\tau_{RC})\approx 7\,{\rm MHz}$.
The amplified voltage is:
\be\label{A(omega)} V_a(\omega)=K(\omega)
V_N=A(\omega)J(\omega),\ee
where $K$ is the amplifier gain and $V_N$ is the input
voltage.

In the long time limit (which is justified since the high
frequency cut-off of the system is much smaller then
$max(eV_s, k_B T)/h$) all current cumulants are
frequency-independent. The amplified voltage fluctuations
$\LL V_a^2\RR=\la [\delta V_a(t)]^2\ra$ bear a simple
relation to the noise power $S=\LL J^2\RR$~\cite{factor2}:
\be  \label{V2} \LL V_a^2\RR=B_2 S, \ \ B_2 =
\frac{1}{2\pi} {\int}_{-\infty}^{+\infty}
A(\omega)A(-\omega)d\omega . \ee
In contrast, the third cumulant $\LL V_a^3\RR=\la [\delta
V_a(t)]^3\ra$, may be decomposed as
\be \label{V3} \LL V_a^3\RR= B_{3}\LL J^3\RR + B_{\rm
en}{\cal J}_{\rm en}+B_{\rm nl}{\cal J}_{\rm nl} \ee
That is, apart the third current cumulant $\LL J^{3}\RR$,
it also contains ``environmental'' cascade correction
${\cal J}_{\rm en}$ originating from the back-action of the
voltage fluctuation across the load on the current
fluctuations in the sample at later times
\cite{KNB02,Nagaev02,Beenakker03,Geneva_group}, and the
correction ${\cal J}_{\rm nl}$ due to nonlinearity of the
sample: 
\bea \label{Venv} &&{\cal J}_{\rm en}=3R_\parallel
S\,dS/dV_N
\\ \label{Vnl}&& {\cal J}_{\rm nl}=
3R^2_\parallel S^2 d^2I/d^2V_N, \eea
where $S=S_s+S_l$ is the noise power generated by the
sample and the load. The coefficients $B_3$, $B_{\rm en}$,
$B_{\rm nl}$, derived below, depend on the circuit only and
are of the same order. The thermal noise of the load and
its resistance are current-independent, and thus do not
contribute to the derivative. Note that since $R_\parallel
S\rightarrow 2k_BT$ at $R_\parallel \rightarrow 0$,
reducing load resistance does not eliminate the
corrections~(\ref{Venv}) and (\ref{Vnl}).

Our measurements are performed using a setup (see
Fig.\,\ref{Setup}) similar to the one discussed in detail
in \cite{Bomze05}. We improved it by increasing the total
gain of the amplification chain to utilize the full 12 bit
resolution of the A/D converter, thus eliminating the
necessity to compensate for its nonlinearity. We also
managed to reduce the nonlinearity of the cryogenic
amplifier by increasing the current through the
transistors. The amplified signal is computer analyzed to
construct the probability distribution function (PDF) of
the amplified voltage~(\ref{A(omega)}). We calibrate the
setup by introducing a known ac signal, spanning the entire
frequency band of the amplifier, through a small capacitor
$C_c\approx 2.4$~pF. This gives us the complex-valued
$A(\omega)$ [see Eq.\ (\ref{A(omega)})] independently at
each value $I$ and $V_g$, which is subsequently used to
compute the coefficients in Eqs.\ (\ref{V2}) and
(\ref{V3}).

To compensate for inaccuracy in the calibration and for the
drift of amplifier parameters we multiply the measured
$A(\omega)$ by a numerical factor, which scales the
measured noise $S$ to the theoretical prediction~\cite{BB}
\be \label{S2}S=eI\left[(1-\Gamma)
\coth\left(\frac{U}{2}\right)+\frac{2\Gamma}{U}\right], \ \
\ U\equiv\frac{eV_s}{k_BT}. \ee
This scale, determined individually for each value of
$V_g$, is found to deviate from unity by less then 10\%.
This is illustrated in the upper inset of Fig.\
\ref{S3+S2}, which shows $S$ fitted by Eq.\,(\ref{S2}) at
$V_g$ corresponding to the
transmission $\Gamma\approx 0.3$ (see the lower inset of
Fig.\ \ref{S3+S2}). We use the scaled $A(\omega)$ to find
the coefficients $B_3$, $B_{\rm en}$, $B_{\rm nl}$, and
then to obtain $\LL J^3\RR$ from Eqs.\ (\ref{V3}) and
(\ref{Venv}). The resulting $\LL J^3\RR$ is shown in the
main panel of Fig.\ \ref{S3+S2}.

As seen in Fig.\,\ref{S3+S2}, the measured third current
cumulant $\LL J^3\RR$ shows very good agreement with the
prediction for noninteracting fermions
\cite{Levitov93,LR01}: \be \label{S3} \LL J^3\RR=e^2I
(1-\Gamma)\left[\frac{6\Gamma}{U}\frac{\sinh(U)-U}{\cosh(U)-1}+1-2\Gamma\right].
\ee
Although the procedure that leads to $\LL J^3\RR$ involves
subtraction of several terms of comparable magnitude, we
stress that it does not rely on any fitting parameter other
than the aforementioned scaling factor. Since we observe no
systematic deviation from the prediction (\ref{S3}), we
believe that the main sources of error in our experiment
are statistical fluctuations (indicated by error bars in
Fig.\ \ref{S3+S2}), as well as our lack of knowledge of the
precise energy dependence of $\Gamma$.

\begin{figure}[tpb]
\includegraphics [angle=-90, width=3.375 in]{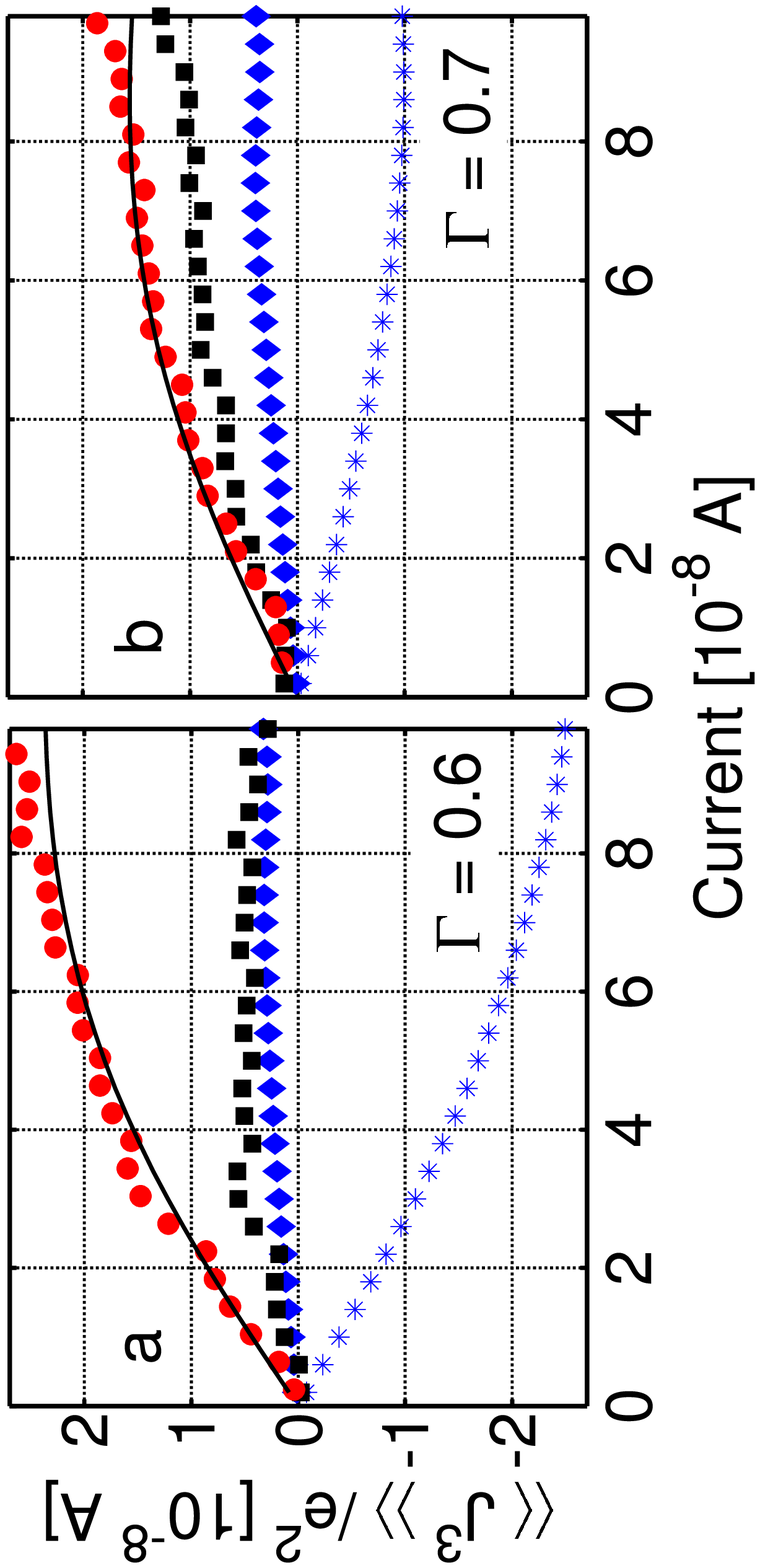} \caption{The contributions used to
evaluate $\LL J^3\RR$ ($\bullet$) from Eqs.\
(\ref{V3}-\ref{Vnl}):
 $\LL V_a^3\RR/B_3 $ ($\blacksquare$),
 ${\cal J}_{\rm en}B_{\rm en}/B_3$ ($\ast$), and
 ${\cal J}_{\rm nl}B_{\rm nl}/B_3$ ($\blacklozenge$).
Panels (a) and (b) show the results obtained with two
different amplifiers, $a$ and $b$, on the sample QPC1 for
similar transmissions $\Gamma=0.6$ and $0.7$. The
coefficients $B_3$, $B_{\rm en}$, and $B_{\rm nl}$ are
calculated using measured response function $A(\omega)$.
Solid line is the prediction (\ref{S3}).} \vspace{-5mm}
\label{TwoSetUps}
\end{figure}

The nearly linear behavior of  $\LL J^3\RR$ at $I\le
10\,{\rm nA}$ corresponds to the low-bias limit of Eq.\
(\ref{S3}):
\be\label{S3l} \LL J^3\RR=e^2 I (1-\Gamma) ,\quad |U|\ll1 .
\ee
The dashed line in Fig.\ \ref{S3+S2} shows the prediction
(\ref{S3l}) with the measured current-dependent
transmission $\Gamma(I)$.
Notably, the full expression (\ref{S3}) provides a much
better fit to our data than the low-bias limit (\ref{S3l}).
In the data taken at larger bias (see Fig.\
\ref{TwoSetUps}), for which the quantity $U$ could be as
large as $4$, the agreement with the expression (\ref{S3})
remains very good. We also note that the large bias limit
of Eq.\ (\ref{S3}), $\LL J^3\RR=e^2 I
(1-\Gamma)(1-2\Gamma)$, exhibits a sign change at
$\Gamma=1/2$. However, the bias regime needed to observe
this effect, $U\gtrsim 10$, is not accessible in our
experiment because of the nonlinearity in $I(V_S)$.

In this work we present the data obtained on two different
samples, QPC1 and QPC2. The sample QPC1 is measured with
two amplifiers, $a$ and $b$, which have different low
frequency cut-offs, $\nu_<^{(a)}\approx 300$\,KHz and
$\nu_<^{(b)}\approx 4$\,MHz, and upper cut-off frequency
$\nu_>\gg 1/(2\pi\tau_{RC})\approx 7\,{\rm MHz}$. For
comparison, we show the results for two similar QPC
transmissions in Fig.\,\ref{TwoSetUps}. Although the
contributions to  $\LL J^3 \RR$ are quite different, the
resulting current cumulants are close and agree well with
the theory. This underscores the necessity of accounting
for the frequency dependent amplifier gain $A(\omega)$ in
data processing. The  low bias results for two different
samples are summarized in the upper inset of
Fig.\,\ref{qpc}: the slope of the $\LL J^3 \RR$ vs.
current, the Fano factor $F_3=\LL J^3
\RR/e^2I$ is proportional to $(1-\Gamma)$, 
confirming universality of the relation\,(\ref{S3l}).

In the rest of the paper we  derive $B_{\rm en}$ and
$B_{\rm nl}$, and express them through $A(\omega)$ and the input circuit parameters. Fluctuations 
are treated in the time domain using the Fourier transforms
$A(t)$, $Z(t)$, and $K(t)$, which vanish for $t<0$ due to
causality. The fluctuation $V(t)=V_N-I R_l$ of the voltage
$V_N(t)$ is generated by the current noise $J$ according to
\be \label{V} C_\parallel\dot V
=-R_\parallel^{-1}V+\frac{d^2I}{dV_N^2}\frac{V^2}{2}+J. \ee
Solving linearized equation (\ref{V}) yields an exponential
relaxation of fluctuations:
\be\label{VN1}
 V(t)=\int dt_1 Z(t-t_1)J(t_1),
\ee
where $Z(t)=C^{-1}_\parallel e^{-t/\tau_{RC}}$. The
amplified voltage $V_a$ is related to $V(t)$ as:
\[
V_a(t)=\int\! dt' K(t-t') V(t')=\int\! dt_1 A(t-t_1)J(t_1)
.
\]

We next ensemble-average the value $[V_a(t)-\langle
V_a\rangle]^3 $. The intrinsic contribution arises from
$\LL J(t_1)J(t_2)J(t_3)\RR$, where $J(t)$ can be treated as
$\delta$-correlated in time. This leads to the first term
in Eq.\ (\ref{V3}) with $B_3$ given by\,\cite{Reulet04}:
\begin{equation}\label{Bint} B_3 = \frac{1}{(2\pi)^2}
{\int\!\!\!\int}_{-\infty}^{+\infty}
A(\omega)A(\omega')A(-\omega-\omega')d\omega d\omega'
\end{equation}
The voltage across the sample fluctuates in time, being a
function of current fluctuations at preceding times, as
described by (\ref{VN1}). As a result, the average $\langle
J(t_1)J(t_2)J(t_3)\rangle$ contains the environmental
contribution due to the dependence of the correlator
\be\label{J2t}\LL  J(t_1)J(t_2)\RR=\delta(t_1-t_2)S(V)\ee
on the voltage $V$, which in turn depends on the current
$J$ at an earlier time $t'$. Linear in $V$ expansion of
Eq.\,(\ref{J2t}) gives the environmental contribution to
Eq.\,(\ref{V3}):
\bea\nonumber 3\int dt_1 A^2(-t_1)\frac{dS}{dV_N} \int
dt'Z(t_1-t') \qquad\qquad\qquad\\ \nonumber \times\LL J(t')
 \int dt_3 A(-t_3) J(t_3)\RR
=3R_\parallel S\frac{dS}{dV_N}B_{\rm en} ,
\\\label{Benv}
B_{\rm en}=R_\parallel^{-1}\int\! dt_1 A^2(-t_1)\int\! dt_3
Z(t_1-t_3)A(-t_3) , \eea 
where the factor 3 accounts for the three possibilities to
choose a later time.

Finally, the nonlinear contribution to Eq.\ (\ref{V3})
comes from the $V^2$ term in Eq.\,(\ref{V}) with the
coefficient
\be \label{Bnl} B_{\rm nl}=R_\parallel^{-2}\int dt
A(-t)\left(\int dt_1 A(-t_1)Z(t-t_1)\right)^2. \ee
For the frequency independent amplification $K$, we find:
\be \label{Ints} B_3=2B_{\rm en}=4B_{\rm
nl}=K^3\tau_{RC}/3{C^3} . \ee
In this case, the first two terms in Eq.\ (\ref{V3}) cancel
at small $I$ and $R_l$. Our amplifier $a$, having
$\nu_<\approx 300\,{\rm KHz}$, operates close to this
regime (see Fig.\ \ref{TwoSetUps}).

It is instructive to note that the ratio of the intrinsic
and environmental coefficients in Eq.\,(\ref{Ints}) is
twice as large as that of Refs.\
\cite{KNB02,Nagaev02,Beenakker03,Geneva_group}. This
difference can be traced to different assumptions about
frequency dependence of the system gain.
Refs.\,\cite{KNB02,Nagaev02, Beenakker03,Geneva_group}
focus on the limit of the high frequency cutoff set by the
amplifier, while in Eq.\,(\ref{Ints}), as well as in our
experiment, it is determined by $Z(\omega)$ with the
roll-off set by $1/\tau_{RC}$. Therefore, our results
correspond to the equal time correlator $\LL
V^3\RR$=$\la[\delta V(t)]^3\ra$.

In summary, we have measured the third cumulant of shot
noise in variable transmission QPCs in the essentially
non-Poissonian regime. In order to extract the
``intrinsic'' third current cumulant we developed a
technique which allows reliable subtraction of the
environmental and nonlinear circuit-dependent
contributions. Good agrement between the experimental
results and the expectations, Ref.\,\cite{Levitov93}, opens
a venue for using high order cumulants as an experimental
tool in mesoscopic physics.

We thank L.~Levitov for valuable discussions,
D.~Goldhaber-Gordon for providing GaAs wafers, and
D.~Shovkun for his contributions at early stages of this
work. This research was supported by the US-Israel
Binational Science Foundation, Israel Science Foundation,
Lady Davis Fellowship, and Swiss NSF.


\vspace{-7mm}


\begin{thebibliography} {qqq}

\bibitem{Glattli}
L. Saminadayar {\em et al.},
%D.\,C. Glattli, Y. Jin and B. Etienne,
% ``Observation of the e/3 fractionally charged Laughlin quasiparticles,''
Phys. Rev. Lett. {\bf 79}, 2526 (1997); R.~de Picciotto
{\em et al.}, Nature (London) {\bf 389}, 162 (1997).

\bibitem{dePicciotto}
M. Reznikov {\em et al.},
%R. de Picciotto, T.\,G. Griffiths, M. Heiblum, and V. Umansky,
Nature (London) {\bf 399}, 238 (1999).
% Observation of a Fifth of the Electron Charge,

\bibitem{Levitov93}
L.S. Levitov, G.B. Lesovik,
% ``Charge distribution in quantum shot noise,''
JETP Lett. {\bf 58}, 230 (1993); cond-mat/9401004

\bibitem{footnote0}
According to Ref.\  \cite{LR01}, it is progressively
difficult to extract high order current cumulants from the
probability distribution of transmitted charge, because the
signal-to-noise ratio scales as $(e/q)^{n/2-1}$.

\bibitem{LR01} L.S. Levitov, M. Reznikov,
% Electron shot noise beyond the second moment
Phys. Rev. B{\bf 70}, 115305 (2004).

\bibitem{Fujisawa06}
T. Fujisawa {\em et al.},
% T. Hayashi, R. Tomita, Y. Hirayama,
% ``Bidirectional Counting of Single Electrons''
Science {\bf 312}, 1634 (2006).

\bibitem{Gustavsson06}  S. Gustavsson {\em et al.},
% Leturcq R, Simovic B, Schleser R, Ihn T, Studerus P, Ensslin K, Driscoll DC, Gossard AC
Phys. Rev. Lett. {\bf 96}, 076605 (2006).

\bibitem{Gustavsson06a}
S. Gustavsson {\em et al.}, Phys. Rev. B 75, 075314 (2007).

\bibitem{ReuletPRL}
B. Reulet, J. Senzier and D.E. Prober, Phys. Rev. Let. {\bf
91}, 196601 (2003).

\bibitem{Bomze05} Yu. Bomze {\em et al.},
% G.\,Gershon, D.\,Shovkun, L.\,S.\,Levitov, and M.\,Reznikov,
Phys. Rev. Lett. 95, 176601 (2005).

\bibitem{Pekola06}
A.V. Timofeev {\em et al.},  Phys. Rev. Lett. {\bf 98},
207001 (2007).

\bibitem{KNB02}
M. Kindermann, Yu.V. Nazarov, C.W.J. Beenakker Phys. Rev.
Lett. {\bf 90}, 246805 (2003).

\bibitem{Nagaev02}
K.E. Nagaev, Phys. Rev. B{\bf 66}, 075334 (2002).

\bibitem{Beenakker03} C.W.J. Beenakker, M. Kindermann, and Yu.V. Nazarov,
Phys. Rev. Let. {\bf 90}, 176802 (2003).

\bibitem{Geneva_group}
S. Pilgram  {\em et al.},
% A. N. Jordan, E. V. Sukhorukov, M. Buttiker,
Phys. Rev. Lett. {\bf 90}, 206801 (2003); A.N. Jordan, E.V.
Sukhorukov, S. Pilgram, J. Math. Phys. {\bf 45}, 4386
(2004).

\bibitem{qpc} B.J. van Wees et al., Phys. Rev. Lett. {\bf 60}, 848 (1988); 
D.A. Wharam et al., J. Phys. C {\bf 21}, L209 (1988).

\bibitem{BB}
Ya. M. Blanter and M. B\"uttiker, Physics Reports {\bf
336}, 1-166 (2000).

\bibitem{factor2}
Our definition of noise moments differs by a factor of $2$
from the one based on the positive frequency representation
for the noise power spectrum,
which brings the Schottky relation to the form $S=eI$.

\bibitem{Reulet04}
B. Reulet {\em et al.}, Proceedings of SPIE, Fluctuations
and Noise in Materials , 5469-33:244-56 (2004);
%% , L. Spietz, C. M. Wilson, G. Senzier, D. E. Prober,
condmat/0403437.





\end{thebibliography}
\end{document}